\documentclass[a4j,10pt,onecolumn,oneside,titlepage,final]{article}

\usepackage{amsmath,amssymb,color,graphicx,amscd,amsfonts,epsf}
\allowdisplaybreaks[4]

\date{}

\begin{document}

\begin{flushright}
KUNS-2120
\end{flushright}

\vspace{0.05cm}

\begin{center}
 {\LARGE
   New inequality for Wilson loops from AdS/CFT
 }
\end{center}
\vspace{0.1cm}
\begin{center}

{\Large          Tomoyoshi Hirata}\\

 \vspace{0.2cm}

E-mail address: hirata@gauge.scphys.kyoto-u.ac.jp

\vspace{0.2cm}

 Department of Physics, Kyoto University,
Kyoto 606-8502, Japan\\
\vspace{0cm}

\end{center}

\vspace{0.5cm}

\begin{center}
 {\bf Abstract}
 \end{center}
 The strong subadditivity is the most important inequality which  entanglement entropy satisfies.
Based on the AdS/CFT conjecture, entanglement entropy in CFT is equal to the area of the minimal surface in AdS space.
It is known that a Wilson loop can also be holographically computed from the minimal surface in AdS space.
In this paper, we argue that Wilson loops also satisfy a similar inequality, and find several evidences of it.
\vspace{0.2cm}
\section{Introduction}

Based on the AdS/CFT correspondence at the large t'Hooft coupling $\lambda$, the expectation value of a Wilson loop $C$ in $D=4$, $\mathit{N}=4$ Super Yang Mills theory is related to the area $A$ of the minimal surface whose boundary is the loop $C$ \cite{Maldacena:1998im}:
\begin{equation}
\langle W(C) \rangle=\exp(- \sqrt \lambda  A).\label{wilson(C)} \quad (\lambda\gg1)
\end{equation}

Recently another object was found to have a connection with minimal surfaces in AdS space.
That is entanglement entropy.
 Based on the AdS/CFT correspondence, when $\lambda\gg 1$, the entanglement entropy of region A in CFT is calculated by replacing the horizon area in the Bekenstein-Hawking formula with the area of the minimal surface in
AdS space whose boundary is the same as that of the region $A$ \cite{Ryu:2006bv} \cite{Ryu:2006ef},
\begin{equation}
 \langle S_A \rangle =\frac{ \ Area(\gamma_ C)}{4G^{(d+2)}_N}  \quad (\lambda\gg1), \label{Entropy=}
\end{equation}
where $G^{(d+2)}_N$ is the Newton constant in $d+2$ dimensional AdS space.

Entanglement entropy always follows a characteristic relation known as  the strong subadditivity \cite{Lieb:1973cp}
\begin{equation}
S_A+S_B \ge S_{A \cup B}+S_{A\cap B}. 
\end{equation}
As Wilson loops and entanglement entropy are related in (\ref{wilson(C)}) and (\ref{Entropy=}) 
through minimal surfaces, Wilson loops should also obey the strong subadditivity at large $\lambda$,
\begin{equation}
\langle W_{\partial A} \rangle  \langle W_{\partial B} \rangle  \ \le\  \langle W_{\partial(A \cup B)} \rangle  \langle W_{\partial(A\cap B)} \rangle.  \label{SS for Wilson}
\end{equation}
Indeed in this paper we point out that the strong subadditivity of Wilson loops is satisfied if we assume minimal surface condition (\ref{wilson(C)}).
We also expect Wilson loops to obey the strong subadditivity in arbitrary coupling regions and find several evidences of it.

This inequality of Wilson loops includes many physical properties, for example, the  convexity of quark potentials, and the convexity of cusp renormalization function.

In this paper, we describe a profound feature of  the strong subadditivity of Wilson loops and study whether or not the strong subadditivity of Wilson loops is satisfied in any coupling region.
To do this, we firstly checked the strong subadditivity in the strong coupling region assuming minimal surface conjecture (\ref{wilson(C)}).
Secondly, usign Bachas inequality we found that the strong subadditivity is satisfied with symmetric Wilson loops with arbitrary coupling constants in any dimensional space. Thirdly, we found that the inequality is satisfied with small-deformed Wilson loops in small coupling regions in any dimension.   

These evidences cause us to conjecture that the strong subadditivity for Wilson loops is satisfied in arbitrary Wilson loops, an arbitrary coupling constant, and an arbitrary dimension. In addition, they give us a criterion of AdS/CFT conjecture (\ref{wilson(C)}).

\section{The strong subadditivity in entanglement entropy and Wilson loops}
 \subsection{Entanglement entropy and its character}

 Consider a quantum mechanical system with many degrees of freedom like a field theory. 
 If the system is put at zero temperature, then the total quantum system is described by the ground state $|\Psi \rangle$. When there is no degeneracy of the ground state, the density matrix is that of the pure state
\begin{equation}
 \rho_{tot}=|\Psi\rangle \langle \Psi|. \label{rho_tot}
\end{equation}

The von Neumann entropy of the total system is clearly zero: $ S_{tot}= -tr\,\rho_{tot} \log \rho_{tot}=0$.
Next we divide the total system into two subsystems, $A$ and $\bar{A}  $.
In the field theory case, we can do this by dividing physical space into two regions and defining $A$ as the field in one region and $\bar{A}$ as the field in the other region
  Notice that physically we do not do anything to the system and the cutting procedure is an imaginary process. Accordingly, the total Hilbert space can be written as a direct product of two spaces
${{H}}_{tot}={{H}}_{A}\otimes {{H}}_{\bar{A}}$ corresponding to those of subsystems $A$ and $\bar{A}$. 

Now we define the reduced density matrix $\rho_A$ by tracing out the Hilbert space
${{H}}_{\bar{A}}$
\begin{equation}
\rho_A=tr_{\bar{A}}\rho_{tot}.
\end{equation} 
 The observer who is only accessible to the subsystem $A$ feels as if the total system were described by the reduced density matrix $\rho_A$.
Because if ${O}_A$ is an operator which acts non-trivially only on $A$, then its expectation value is
\begin{equation}
O_A=tr{O}_A\cdot \rho_{tot}=tr_{A}{ O}_A\cdot \rho_{A}
\end{equation}
where the trace $tr_A$ is taken only over the Hilbert space ${H}_{A}$.

Then we define entanglement entropy of the subsystem $A$ as the von Neumann entropy of the reduced density matrix $\rho_A$
\begin{equation}
{S_A= - tr_{A}\, \rho_{A} \log \rho_{A}.}
\end{equation}
 This entropy measures the amount of information lost by tracing out the subsystem $\bar{A}$.
One can define entanglement entropy by choosing another total density matrix than (\ref{rho_tot}). However, this choice is sufficient for the purposes of this paper. 

Entanglement entropy satisfies many inequalities: the most important one being the the strong subadditivity 
\begin{equation}
S_A+S_B \ge S_{A \cup B}+S_{A\cap B}. \label{SS for EE}
\end{equation}
This inequality is also satisfied by any general density matrix.
The strong subadditivity is the strongest inequality of the von-Neumann entropy. Indeed, it has mathematically been shown that the strong subadditivity in conjunction with several other more obvious conditions (such as the invariance under unitary transformations and the continuity with respect to the eigenvalues of $\rho_{tot}$) characterize the von-Neumann entropy \cite{Aczel etal}.

According to AdS/CFT correspondence, any physical quantity of $d+1$ dimensional CFT theory can be gained from the dual $d+2$ dimensional anti de-Sitter space ($AdS_{d+2}$).
This is also the case with entanglement entropy.
In \cite{Ryu:2006bv} and \cite{Ryu:2006ef}, it is clamed that entanglement entropy of $d$ dimensional spacelike submanifold $A$ in $d+1$ dimensional CFT theory is given by the following formula:
\begin{equation}
S(A)=\frac{ \ Area(\gamma_ A)}{4G^{(d+2)}_N}. \label{S(A)=}
\end{equation}
 where $Area(\gamma_A)$ denotes the area of the surface $\gamma_A$, and $G^{(d+2)}_N$ is the Newton constant in the $d+2$ dimensional anti de-Sitter space. The
$d$ dimensional surface $\gamma_A$ is defined as the surface with minimal area whose boundary coincides with the boundary of submanifold $A$.

The conjecture (\ref{S(A)=}) is mathematically proved in two dimensional CFT and 
in general dimensional case a good explanation is given in \cite{Fursaev:2006ih}.

In addition, 
 \cite{Hirata:2006jx} shows a numerical evidence to prove that the holographic entanglement entropy (\ref{S(A)=}) follows the strong subadditivity, and \cite{Headrick:2007km} gives a mathematical proof of it.

 \subsection{The strong subadditivity of Wilson loops}\label{The strong subadditivity for Wilson loops}
 
From (\ref {wilson(C)}) and (\ref{Entropy=}), the strong subadditivity of  entanglement entropy (\ref{SS for EE}) is translated into that of Wilson loop
\begin{equation}
\langle W_{\partial A} \rangle  \langle W_{\partial B} \rangle  \ \le\  \langle W_{\partial(A \cup B)} \rangle  \langle W_{\partial(A\cap B)} \rangle, 
\end{equation}
where $W_{\partial A}$ is a Wilson loop around the boundary of region $A$. 
To determine regions $A$ and $B$ for Wilson loops, we only consider the case where Wilson loops are in spacelike two-dimensional flat plane.

If the left hand side and the right hand side of (\ref{SS for Wilson}) are not real, then the inequality becomes meaningless. However, when the system is invariant under charge conjugation $A_\mu \to -A_\mu^T$, the value of a Wilson loop in Euclidian space is real because when subjected to the charge conjugation we have
\begin{eqnarray}
\langle e^{ P i \oint dx^\mu i g A_\mu  }\rangle  \to \langle e^{P i \oint dx^\mu i (-g A_\mu^T)  } \rangle 
= \langle e^ {P i \oint dx^\mu i g A_\mu  } \rangle ^*.
\end{eqnarray}

We also note that if Wilson loops follow only the area law and the perimeter law,
\begin{equation}
\langle W_{\partial A} \rangle= \exp(-K_1  S(A) - K_2 L( \partial A) ),\label{area and perimeter low}
\end{equation}
where $K_1$ and $K_2$ are constants and $S(A)$ is the area of $A$  and $L(A)$ is the length of the perimeter, the expectation value of Wilson loops follows the equality,
 \begin{equation}
\langle W_{\partial A} \rangle  \langle W_{\partial B} \rangle  \ = \  \langle W_{\partial(A \cup B)} \rangle  \langle W_{\partial(A\cap B)} \rangle.  \label{SE for Wilson}
\end{equation}
This is because
\begin{eqnarray}
 S(A) +S(B)&=& S(A \cup B) + S(A\cap B) \\
 L(A) +L(B)&=& L(A \cup B) + L(A\cap B) .
\end{eqnarray}
Therefore, the subadditivity comes from other factors than the area and the perimeter law factor.
One interesting example of the equality (\ref{SE for Wilson}) is in the large $N$ pure Yang-Mills lattice QCD${}_2$. There, loop equations are easily solved \cite{Paffuti:1980cs}\cite{Makeenko:2002uj}, and nonintersecting Wilson loops are calculated to
\begin{eqnarray}
W(C)&=&\left(1- \frac{\lambda a^2}{2}\right)^{A/a^2} \ (\lambda < 1)\\
W(C)&=&\left(\frac{1}{2\lambda a^2}\right)^{A/a^2} \ \ \ \ (\lambda >1),
\end{eqnarray}
where $a$ is the lattice spacing. Therefore, Wilson loops follow the pure area law both in weak coupling and in strong coupling regions. Hence, from the argument above, Wilson loops satisfy the equality (\ref{SE for Wilson}).

In the next section, we will show three important applications of the strong subadditivity for Wilson loops.

\section{Application of the strong subadditivity for Wilson loops}\label{ASSWL}
\subsection{Cusp anomalous dimensions} \label{subsection Cusp}
In this subsection, we consider the renormalization of Wilson loops in four-dimensional Yang-Mills theory.
When a Wilson loop has cusps whose angles are $\theta_k$ respectively, the renormalization of Wilson loops is multiplicatively renormalizable \cite{Brandt:1981kf}\cite{Polyakov:1980ca},such that:

\begin{equation}
\langle W^{ren}(M,C) \rangle= Z_{per}(M,C) \prod^k Z_{cusp}(M,\theta_k) \langle W^{non\ ren}(C)\rangle,
\end{equation}
where $M$ is a renormalization scale, and $Z_{per}$ is renormalization constant which comes from perimeter, $Z_{cusp}(M,\theta_k)$ is an additional renormalization constant which comes from cusps, and $W^{ren}$ is finite when expressed via the renormalized charge.

By solving the Callan-Symanzik equation for Wilson loop \cite{Korchemsky:1987wg} we have
\begin{equation}
Z_{cusp}(M,\gamma)=g_R(M)^{-\Gamma_{cusp}(\gamma)/ C_\beta},
\end{equation}
where $\Gamma_{cusp}(\gamma)$ is an anomalous dimension of $Z_{cusp}(\gamma)$ and called a "cusp anomalous dimension", $g_R(M)$ is the renormalized coupling constant, and $C_\beta$ is the coefficient of the $\beta$ function:
\begin{equation}
C_\beta= \frac{11}{3} N_c -\frac{2}{3} N_f.
\end{equation}

\begin{figure}[htbp]
\begin{tabular}{ccc}
\begin{minipage}[t]{0.45\hsize}
\begin{center}
 \includegraphics[scale=0.5]{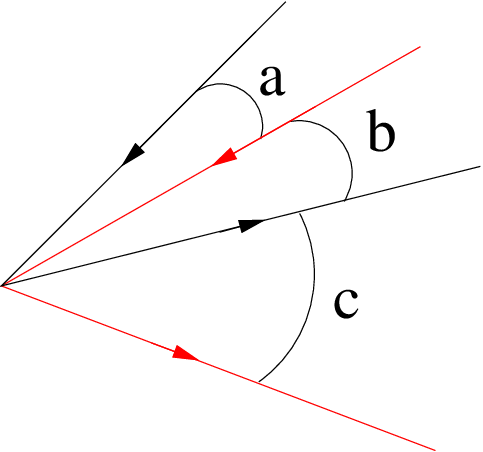}
  \caption{Two Wilson loops with cusp. One's angle is $a+b$ and the other's is $b+c$. }
\label{fig:cusp.eps}
\end{center}
\end{minipage}
\begin{minipage}{0.1\hsize}
\end{minipage}
\begin{minipage}[t]{0.45\hsize}
\begin{center}
 \includegraphics[scale=0.5]{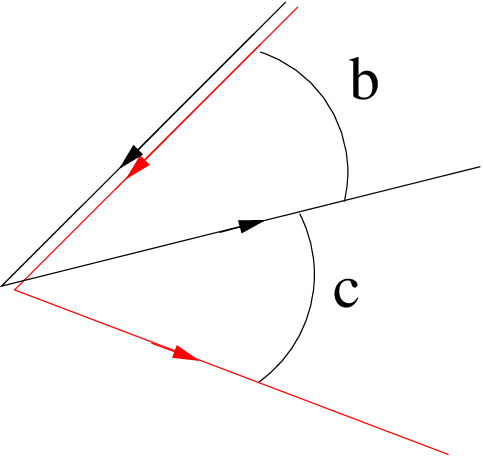}
  \caption{When all crossing points of two loops are like this, the strong subadditivity gives a stronger condition than (\ref{conv}). }
 \label{fig:touch.eps}
\end{center}
\end{minipage}
\end{tabular}
\end{figure}

Now we apply the strong subadditivity of $\langle W^{non\ ren}(C)\rangle $ to two Wilson loops whose cusp angles are $a+b$ and $b+c$ respectively (Fig.\ref{fig:cusp.eps}).

Up to the second order perturbation $\Gamma_{cusp}<0$. So when $M\gg 1$ and $g_R(M)\ll 1$, $Z_{cusp}^{-1}(M,\gamma)$ is much greater than  $1$.
$Z_{per}$ should cancel each other on both sides of the inequality, as divergence of entanglement entropy derived from perimeters cancel each other out. Then we  have 
\begin{equation}
\Gamma_{cusp}(a + b + c)+\Gamma_{cusp}( b)  \le  \Gamma_{cusp}(a+b )+\Gamma_{cusp}( b+c ) \label{conv}
\end{equation}

This leads to the convexity
\begin{equation}
\frac{d^2}{d^2 \theta} \Gamma_{cusp}(\theta)\le 0.
\end{equation}

Up to the second order perturbation \cite{Korchemsky:1987wg} we have
\begin{equation}
\Gamma_{cusp}^{(2)}(\theta)= 4 \frac{N_c^2-1}{2N_c}(\theta\cot \theta -1)\quad (<0)
\end{equation}
for $SU(N_c)$ gauge theory.
Then we can directly check the convexity of $ \Gamma_{cusp}^{(2)}(\theta)$ as
\begin{equation}
\frac{d^2}{d^2 \theta} \Gamma^{(2)}_{cusp}(\theta)=4 \frac{N_c^2-1}{2N_c}(\theta\cot \theta -1) \frac{2}{\sin^2(\theta)}<0.
\end{equation}

Conversely, (\ref{conv}) derives the strong subadditivity when two loops have a crossing point like Fig.\ref{fig:cusp.eps}, because ${Z_{cusp}}$ are  main divergence parts of  loops with  cusps.
(\ref{conv}) saturates when $a=0$ or $b=0$.
Therefore when all crossing points of two loops  are like Fig.\ref{fig:touch.eps} (i.e. when two loops don't cross but just touch),\footnote{A good example is shown in the next subsection (Fig.\ref{fig:quark.eps}).
} the strong subadditivity gives a nontrivial condition except (\ref{conv}).\footnote{A similar situation also occurs in the case of Bachas inequality \cite{Pobylitsa:2007ky}}

\subsection{Quark potential} \label{subsection quark}
\begin{figure}[htbp]
  \begin{center}
    \includegraphics[scale=0.5]{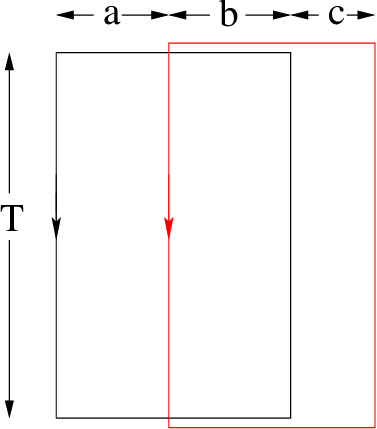}
  \end{center}
  \caption{Rectangular Wilson loops}
  \label{fig:quark.eps}
\end{figure}

We consider rectangular Wilson loops shown in Fig.\ref{fig:quark.eps} where short sides of the rectangle are $a+b$ and $b+c$, and long sides of them are all $T$.
 
The value of Wilson loops $W(R,T)$ has a physical meaning as a quark potential $V(R)$:
\begin{equation}
V(R) =- \lim_{T\to \infty}\ln W(R,T).
\end{equation}

The strong subadditivity of these Wilson loops is
\begin{equation}
V(a+b+c)+V(b) \le V (a+b) + V(b+c)
\end{equation}
This is equivalent with the convexity of quark potential
\begin{equation}
\frac{d^2}{d^2 R} V(R)\le 0.
\end{equation}

As shown in \cite{Pobylitsa:2007ky}\cite{Bachas:1985xs} , one can also derive the convexity of cusp anomalous dimensions and the convexity of quark potential from Bachas inequality, which we will consider in the next section.
\subsection{Inequality of $F_{\mu \nu}$ inserted Wilson loop} \label{subsection:F}
\begin{figure}[htbp]
  \begin{center}
    \includegraphics[scale=0.5]{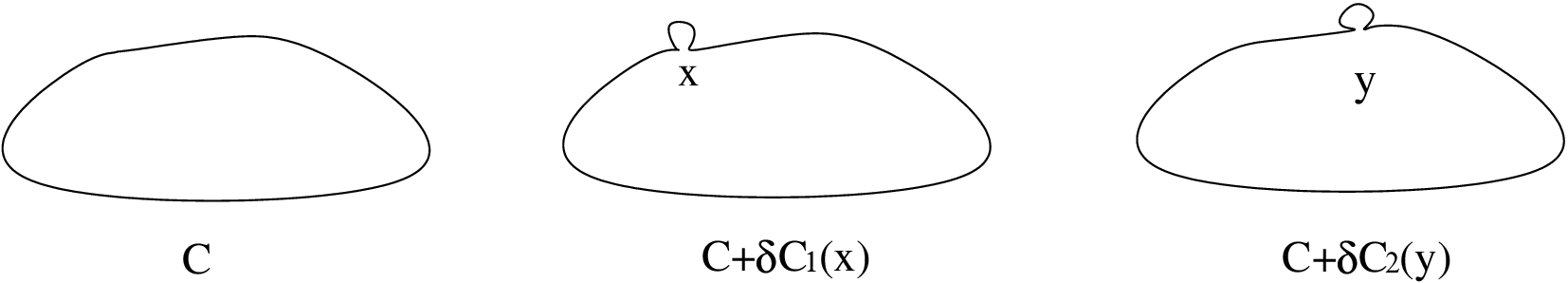}
  \end{center}
  \caption{Wilson loop C and its small deformed Wilson loops, $C+\delta C_1(x)$ and $C+\delta C_2(y)$. }  \label{fig:CdeltaC.eps}
\end{figure}

Let us consider three loops $C$, $C+\delta C_1(x)$ and $C+\delta C_2(y)$, where $\delta C_1(x)$ and 
$\delta C_2(y)$  are  infinitesimal loops attached to a loop $C$ at points $x$ and $y$ ($x \neq y$) respectively and are located outside of $C$ (Fig.\ref{fig:CdeltaC.eps}).

 Now we can see
\begin{eqnarray}
Int(C+\delta C_1) \cup Int(C+\delta C_2)& =& Int(C+\delta C_1+\delta C_2)\\
Int(C+\delta C_1) \cap Int(C+\delta C_2) &=& Int(C),
\end{eqnarray}
where $Int(C)$ means the interior of $C$.
The strong subadditivity for $C+\delta C_1$ and $C+\delta C_2$ is 
\begin{equation}
\langle W(C+ \delta C_1) \rangle \langle W(C+ \delta C_2) \rangle \ \le\ \langle W(C+ \delta C_1 +\delta C_2)\rangle \langle W(C) \rangle \label{area deformation SS for Wilson}.
\end{equation}

Using area derivative, we can expand $W(C+ \delta C)$ as
\begin{equation}
W(C+ \delta C)=W(C) + \delta \sigma^{\mu \nu} \frac{\delta W(C)}{\delta \sigma^{\mu \nu}(x)} 
+ \frac{1}{2} \left(\delta \sigma^{\mu \nu} \frac{\delta W(C)}{\delta \sigma^{\mu \nu}(x)}\right) ^2 +O((\delta\sigma^{\mu \nu})^3),
\end{equation}
where $\delta \sigma^{\mu \nu}$ denotes the area enclosed by $\delta C^{\mu \nu}$.

The area derivative is given by inserting the field strength $i F_{\mu \nu}$ into Wilson loop:
\begin{equation}
 \frac{\delta }{\delta \sigma^{\mu \nu}(x)} W(C)= tr \mathbf{P}( iF_{\mu \nu}(x) e^{i \oint _c d \xi ^\alpha A_{\alpha}}),
 \end{equation}
 therefore the inequality (\ref{area deformation SS for Wilson}) can be rewritten as 
\begin{equation}
0 \ge \langle W(C) \rangle \langle tr \mathbf{P}( (F\cdot \delta\sigma)_x
 (F\cdot \delta\sigma)_y 
 e^{i \oint _c d \xi ^\mu A_{\mu}}) \rangle, \label{FF le 0}
\end{equation}
where $(F\cdot \delta\sigma)_x$ denotes $F_{\mu \nu}(x) \delta\sigma^{\mu \nu}$.
  
This inequality is  fundamental for the strong subadditivity.
Indeed all inequalities of the strong subadditivity of small-deformed Wilson loops are derived from (\ref{FF le 0}).
Firstly let us consider three loops $C$, $C+\sum _i \delta C_i(x_i)$ and $C+\sum \delta C_j(y_j)$, where $\delta C_j(x_j)$ and 
$\delta C_i(y_i)$  are  infinitesimal loops attaced to a loop $C$ at points $x_i$ and $y_j$ ($x_i$ and $y_j$ are all different points) respectively and are located outside of $C$. 
The reason why we don't have to consider a case where $\delta C_i(x_i) $ or  $\delta C_j(y_j) $ are inside $C$ 
or a case where $x_i$ and $y_j$ are not all different points is@that in those cases by redefining $C$ as $(C+\sum _i \delta C_i(x_i)) \ \cap (\ C+\sum \delta C_j(y_j))$ we can regain the original situation. 
 
 Now the strong subadditivity is
\begin{equation}
    \langle W(C+\sum _i \delta C_i(x_i))  \rangle \langle W( C+\sum \delta C_j(y_j)) \rangle 
 \le \langle W(C) \rangle \langle W( C+\sum _i \delta C_i(x_i)+\sum \delta C_j(y_j)) \rangle . \label{general C_i}
\end{equation}
Rewriting it by the operator form, (\ref{general C_i}) is
\begin{equation}
0 \ge \sum _{i,j}\langle W(C) \rangle  \langle tr \mathbf{P}( (F\cdot \delta\sigma)_{x_i} (F\cdot \delta\sigma)_{y_j} e^{i \oint _c d \xi ^\mu A_{\mu}} )\rangle. \label{FFs le 0}
\end{equation}
We can obtain this inequality from the former inequality (\ref{FF le 0}). Other kind of small deformed Wilson loops are obtained by summing infinite number of $\delta C_i$. Therefore, the inequality (\ref{FF le 0}) will be the most essential inequality for the strong subadditivity of Wilson loops. 
In the next section, we will give a perturbative proof of the strong subadditivity for general small-deformed Wilson loops, which gives a proof of (\ref{FF le 0}) as a special case.
\section{Verification of the strong subadditivity of Wilson loops}
In this section, we verify  the strong subadditivity of Wilson loops in three ways.

Firstly, we assume AdS/CFT conjecture (\ref{wilson(C)}) and from the nature of the minimal surface we prove the inequality at $\lambda \gg 1$. Secondly, using Bachas inequality, which specially-shaped Wilson loops satisfy,
we prove the strong subadditivity of specially-shaped Wilson loops in all coupling regions. 
Thirdly, we give a perturbative proof of the strong subadditivity for all small-deformed Wilson loops.

\subsection{Verification from the minimal surface conjecture} \label{sec minimal}
Assuming AdS/CFT conjecture (\ref{S(A)=}) in the strong coupling region, it is possible to prove the strong subadditivity (\ref{SS for Wilson}).
We use the same logic as the proof for the strong subadditivity of entanglement entropy shown in \cite{Headrick:2007km}.

\begin{figure}[htbp]
  \begin{center}
    \includegraphics[scale=0.5]{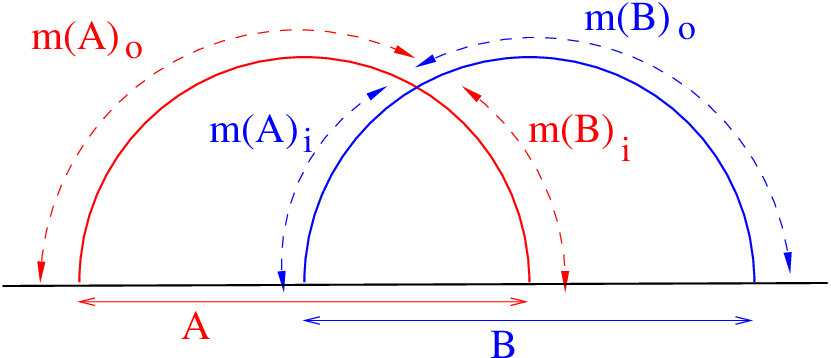}
  \end{center}
  \caption{region $A$, $B$ and their minimal surfaces}
  \label{fig:minimal}
\end{figure}

Let $m(A)$ and $m(B)$  be the minimal surface of region A and B respectively where A and B are interiors of Wilson loops.
$m(A)$ is divided by $m(B)$.
Let $m(A)_o$ be outside piece of $m(A)$ with respect to $m(B)$ and
let $m(A)_i$ be inside piece of $m(A)$ with respect to $m(B)$.
We also define $m (B)_o$, $m(B)_i$ in the same way (Fig.\ref{fig:minimal}).

Since $m(A)_o \cup m(B)_o$ is a surface whose boundary is  $A \cup B$, its area is bigger than or equal to the area of the minimal surface $m(A \cup B)$ whose boundary is $A \cup B$:
\begin{equation}
m(A)_o + m(B)_o \ge m(A \cup B).
\end{equation}
And more since $m (A)_i \cup m(B)_i$ is a surface whose boundary is  $A \cap B$, its area is bigger than or equal to the area of the minimal surface $m(A \cap B)$ whose boundary is $A \cap B$:
\begin{equation}
m(A)_i + m(B)_i \ge m(A \cap B)
\end{equation}

Therefore, we have an inequality 
\begin{equation}
m(A)+m(B)=m(A_o)+ m(A_i)+m(B_o) +m(B_i) 
\ge  m(A \cup B)  +m(A \cap B). \label{area inequality}
\end{equation}
The strong subadditivity can be derived from this.

One note should be mentioned for this subsection.
Here, we have considered the case where two Wilson loops are on a same flat plane.
However, when two loops are on a same curved non-intersecting surface, we can also prove the strong subadditivity in the same way.

\subsection{Verification from Bachas inequality}
\subsubsection{Review of Bachas inequality}
Here we review Bachas inequality.
We define $\theta$ as Parity transformation along $x^1$ axis and region $L_0$ $L_+$ $L_-$ as
\begin{eqnarray}
L_0&\equiv&\{x^\mu  ;x^1=0\}\\
L_+&\equiv&\{x^\mu  ;x^1>0\}\\
L_-&\equiv&\{x^\mu  ;x^1<0\}.
\end{eqnarray}
Now let $C_i$ be open lines which exist in $L_+$ and let their boundaries in $L_0$.
Now we define a function $f_{ab}$ as
 \begin{equation}
f_{ab}= \sum_{i}^{M} k^i W(C_i)_{ab}, \label{f_{ab}}
\end{equation}
where $ W(C_i)_{ab}$ is a Wilson line operator of $C_i$ , $k_i$ is an arbitrary real number, and $a$ $b$ are gauge indices.

Then we have
\begin{eqnarray}
tr \langle f \theta f^\dagger \rangle 
&=&Z^{-1} tr \int \prod _{b\in L_0}dU_b e^{-S_0}
\int \prod _{b\in L_+} dU_b f(U(b)) e^{-S_+}  \nonumber
 \int \prod _{b\in L_-} dU_b f(U(\theta b)) ^{\dagger} e^{-S_-} \\
&=&Z^{-1} \int \prod _{b\in L_0}dU_b e^{-S_0}
\left|\int \prod _{b\in L_+} dU_b f(U(b)) e^{-S_+}\right|^2 \ge 0 \label{ftf}
\end{eqnarray}
where $S_0$, $S_+$ and  $S_-$ are respectively actions in $L_0$, $L_+$ and $L_-$.
Substituting (\ref{f_{ab}}) we have
\begin{equation}
tr \langle f \theta f^\dagger \rangle =\sum_{ij} k^i \langle W(C_{ij}) \rangle k^j,
\end{equation}
where $W(C_{ij})$ is the Wilson loop made by $C_i$ and mirror image of $C_j$ 
\begin{equation}
\langle W(C_{ij}) \rangle=tr\langle W(C_i) W(-\theta C_j) \rangle \label{quadratic}.
\end{equation}
Therefore the inequality (\ref{ftf}) means the quadratic form (\ref{quadratic})
is positive definite i.e. the determinant of the matrix $\langle  W(C_{ij})  \rangle $ is positive.
\begin{equation}
\det_{ij} \langle W(C_{ij}) \rangle\ge 0 \label{Bachas ineq}
\end{equation}

This is Bachas inequality which was extended by Pobyltsa \cite{Pobylitsa:2007ky}.
When $M=2$ (\ref{Bachas ineq}) derives the original inequality presented by Bachas \cite{Bachas:1985xs},
\begin{equation}
\langle W(C_{11})\rangle \langle W(C_{22}) \rangle \ge \langle W(C_{12})\rangle \langle W(C_{21})\rangle \label{original BI}
\end{equation}

\subsubsection{Bachas inequality and the strong subadditivity: first example}
Now we will see some Bachas inequalities are equivalent to or are derived from the strong subadditivity for some symmetric Wilson loops but in all coupling region. 

\begin{figure}[htbp]
\begin{tabular}{ccc}
\begin{minipage}[t]{0.45\hsize}
\begin{center}
 \includegraphics[scale=0.5]{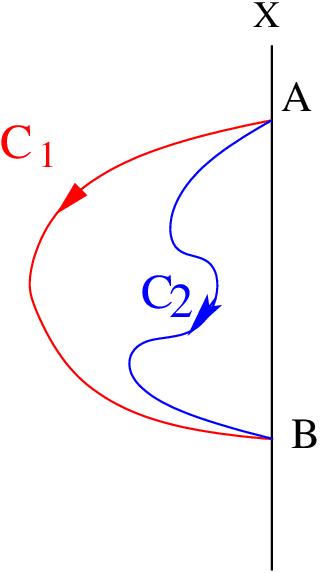}
\caption{Two open lines $C_1$ and $C_2$ and axis $X$.
 The interior of $C_1 + X$ is outside of the interior of $C_2 +X$.}
\label{fig:bachas1.eps}
\end{center}
\end{minipage}
\begin{minipage}{0.1\hsize}
\end{minipage}
\begin{minipage}[t]{0.45\hsize}
\begin{center}
 \includegraphics[scale=0.5]{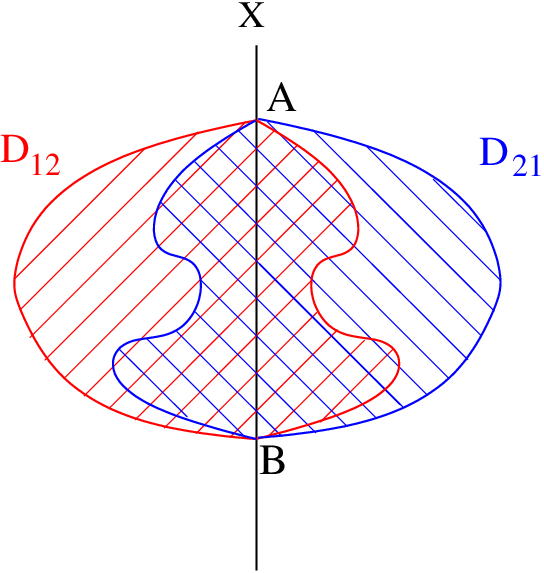}
\caption{Configuration of $D_{12}$ and  $D_{21}$.}
 \label{fig:121st.eps}
\end{center}
\end{minipage}
\end{tabular}
\end{figure}

Consider two open lines $C_1$ and $C_2$ which touch an axis $X$.
Let \begin{equation}
Int(C_1+X) \supset Int(C_2+X),
\end{equation}
where $Int(C_i+X)$ is the interior of $C_i+X$.
We consider the case where each $C_1$'s two end points A and B are at the same place as $C_2$'s end points (Fig.\ref{fig:bachas1.eps}).

 As can be seen from Fig.\ref{fig:121st.eps},
\begin{eqnarray*}
D_{12} \cup  D_{21} = D_{11}, \quad D_{12} \cap  D_{21} = D_{22}
\end{eqnarray*} 
are satisfied where $D_{ij}$ is the interior of $C_{ij}$.

So now the original Bachas inequality 
\begin{equation}
 \langle W(C_{11})\rangle\langle W(C_{22})\rangle\ge \langle W(C_{12})\rangle\langle W(C_{21})\rangle,
\end{equation}
 is equivalent to the strong subadditivity 
 \begin{equation}
\langle W(\partial( D_{12} \cap  D_{21}) )\rangle\langle W(\partial(D_{12} \cup  D_{21}))\rangle\ge \langle W(\partial D_{12})\rangle\langle W(\partial D_{21})\rangle.
\end{equation}
 
 This example includes examples shown in section \ref{subsection Cusp} and section \ref{subsection quark} \cite{Pobylitsa:2007ky}\cite{Bachas:1985xs}.
 \subsubsection{Bachas Inequality and the strong subadditivity: second example}
\begin{figure}[htbp]
  \begin{center}
    \includegraphics[scale=0.5]{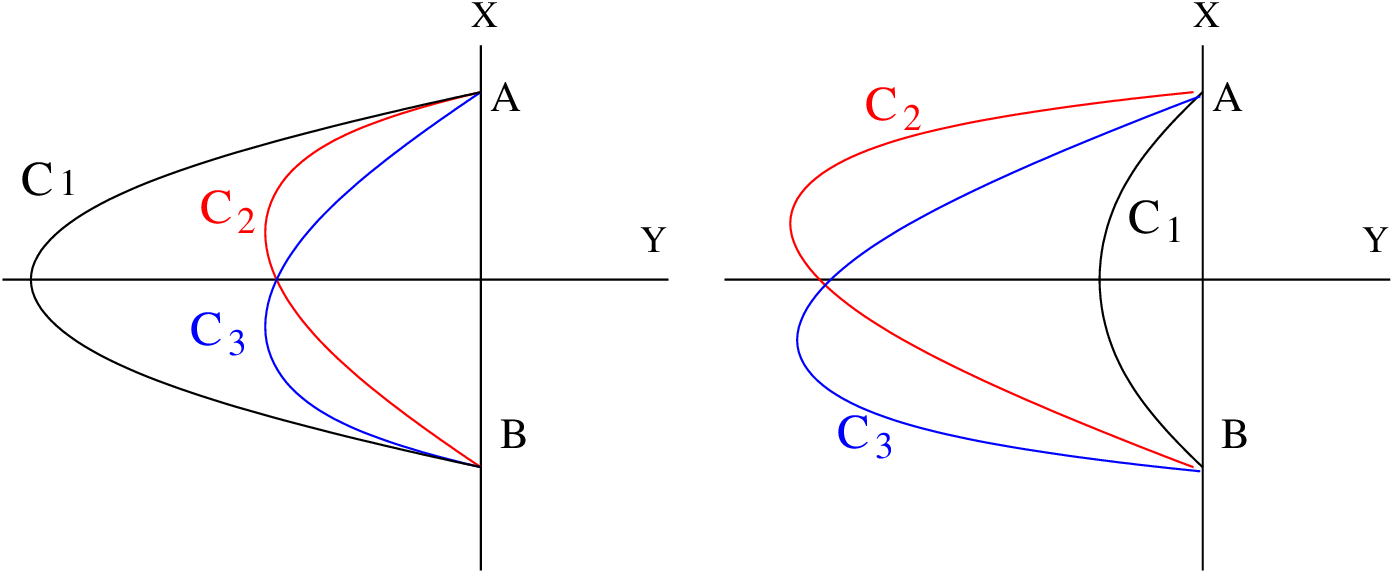}
  \end{center}
  \caption{Three open lines: $C_1$,$C_2$ and $C_3$. The interior of $C_1 + X$ is inside or outside of the interior of $C_2 +X$, $C_3 +X$. $C_2$ and $C_3$ are symmetric with respect to the Y-axis, $C_1$ is axisymmetric with respect to the Y-axis.}
  \label{fig:bacas3.eps}
\end{figure}

Let us introduce three open lines $C_1$,$C_2$ and $C_3$ which touches X-axis.
Now we impose following conditions to these lines.
Firstly the interior of $C_1 + X$ is inside or outside of the interior of $C_2 +X$, $C_3 +X$:
\begin{equation}
Int (C_1+X) \supset Int(C_2 +X),\ Int(C_3 +X)
\end{equation}
or
\begin{equation}
Int (C_1+X) \subset Int(C_2 +X),\ Int(C_3 +X).
\end{equation}
Secondly
$C_2$ and $C_3$ are symmetric with respect to the Y-axis which is perpendicular to X axis and $C_1$ is axisymmetric with respect to the Y-axis. 
Thirdly $C_1$, $C_2$ and $C_3$ share their two end points A and B (Fig.\ref{fig:bacas3.eps}).\footnote{Originally the Bachas equation for this configuration was considered in \cite{Pobylitsa:2007ky}}

From their symmetry, we have
\begin{eqnarray}
W(C_{12})&=&W(C_{21})=W(C_{13})=W(C_{31})\\
W(C_{22})&=&W(C_{33}).
\end{eqnarray}
Now let us define $r,t,x,p$ as
\begin{eqnarray}
&&r=\langle W(C_{11}) \rangle ,\ x=\langle W(C_{12}) \rangle =\langle W(C_{13}) \rangle \nonumber \\
&&t=\langle W(C_{22}) \rangle =\langle W(C_{33}) \rangle ,\ p=\langle W(C_{23}) \rangle 
\end{eqnarray}
Then Bachas inequalities (\ref{Bachas ineq}) for these three paths lead to 
\begin{eqnarray}
\det_{i,j=2}\langle W(C_{ij}) \rangle&=& {t}=t \ge 0 \label{t ge 0}\\
\det_{i,j=2,3}\langle W(C_{ij}) \rangle
&=&
\det{\left(
  \begin{array}{cc}
    t   &p    \\
     p&t    \\
  \end{array}
\right)
} = t^2 -p^2 \ge 0 \label{r^2-p^2 ge 0}\\
\det_{i,j=1,2,3}\langle W(C_{ij}) \rangle&=&\det{\left(
  \begin{array}{ccc}
    r   & x   & x   \\
     x  & t   &p    \\
     x  &   p &   t \\
  \end{array}
\right)
}=(t-p) [r(t+p) -2x^2] \ge 0 \label{det ge 0}
\end{eqnarray}
(\ref{t ge 0})(\ref{r^2-p^2 ge 0}) leads to $t\ge p$. Therefore if $t \neq p$, (\ref{det ge 0}) results in 
\begin{equation}
r(t+p) \ge 2x^2. \label{r(t+p) ge}
\end{equation}
This inequality also holds if $t=p$, because in this case (\ref{r(t+p) ge}) is equivalent to another Bachas inequality
\begin{eqnarray}
\det_{i,j=1,2}\langle W(C_{ij}) \rangle=\det{\left(
  \begin{array}{cc}
    r   &x    \\
     x&t    \\
  \end{array}
\right)
} = rt -x^2&\ge& 0.
\end{eqnarray}

\begin{figure}[ht]
  \begin{center}
    \includegraphics[scale=0.4]{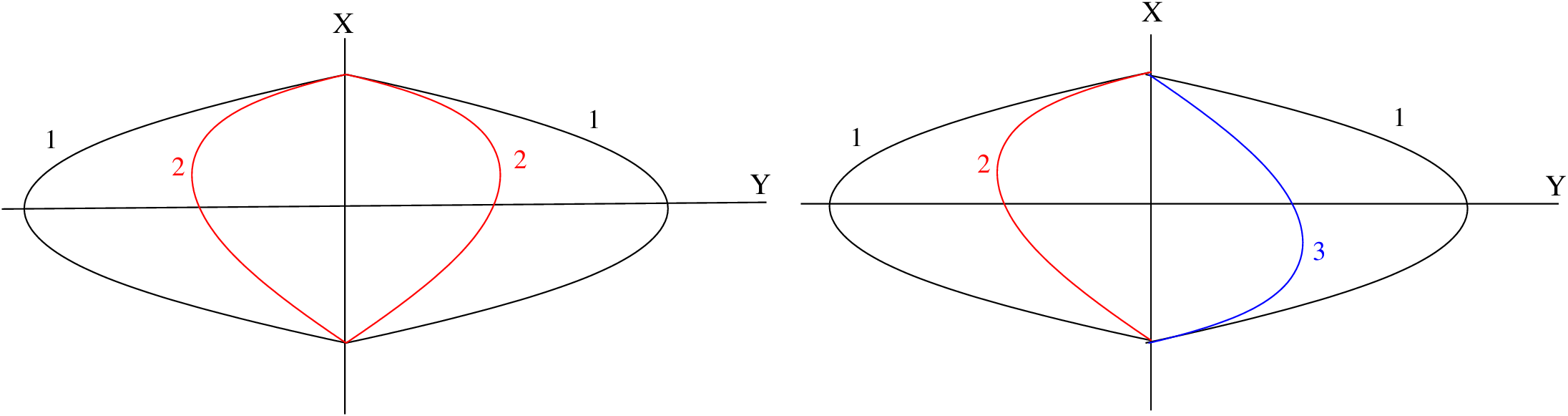}
  \end{center} \caption{
These two figures show the configuration of $C_1$, $C_2$, $C_3$ and their mirror images when the interior of $C_1 + X$ is inside of the interior of $C_2 +X$ and $C_3 +X$. The figure on the left shows (\ref{D12}), while that on the right shows (\ref{D123}). 
  }
  \label{fig:d12s.eps}
\end{figure}

\begin{figure}[hbtp]
  \begin{center}
    \includegraphics[scale=0.4]{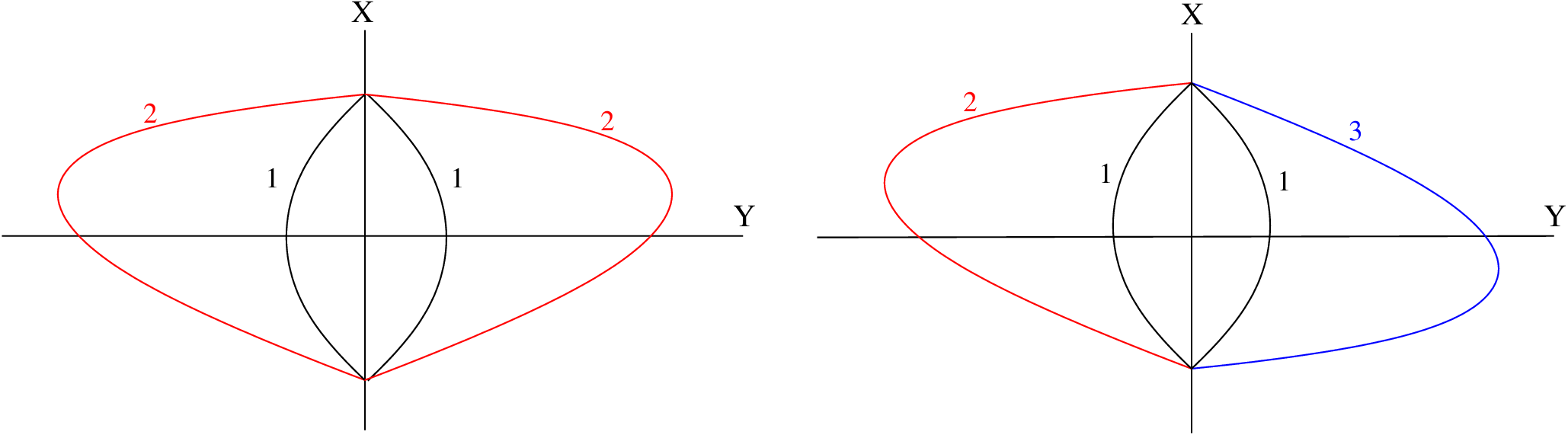}
  \end{center}
  \caption{
 These two figures show the configuration of $C_1$, $C_2$, $C_3$ and their mirror images when the interior of $C_1 + X$ is outside the interior of $C_2 +X$ and $C_3 +X$. The figure on the left shows (\ref{D12"}), while that on the right shows (\ref{D123"}). }
  \label{fig:d12s2nd.eps}
\end{figure}

On the other hand, one can also derive (\ref{r(t+p) ge}) by the strong subadditivity.
If the interior of $C_1 + X$ is inside of the interior of $C_2 +X$, we obtain 
\begin{eqnarray}
D_{12} \cap D_{21} = D_{11},\quad D_{12} \cup D_{21} = D_{22} \label{D12}\\ 
D_{13} \cap D_{21} = D_{11},\quad D_{13} \cup D_{21} = D_{23} \label{D123}
\end{eqnarray}
(See Fig.\ref {fig:d12s.eps}). 
If the interior of $C_1 + X$ is outside of the interior of $C_2 +X$, we obtain 
\begin{eqnarray}
D_{12} \cup D_{21} = D_{11}, \quad D_{12} \cap D_{21} = D_{22} \label{D12"}\\
D_{13} \cup D_{21} = D_{11}, \quad D_{13} \cap D_{21} = D_{23} \label{D123"}
\end{eqnarray}
(see Fig.\ref{fig:d12s2nd.eps}) .

In each case the strong subadditivity gives
\begin{eqnarray}
x^2\le rt \quad x^2 \le rp. \label{x^2le rp}
\end{eqnarray}
These inequalities lead to 
\begin{equation}
r(t+p) \ge 2x^2
\end{equation}
which is the same inequality as (\ref{r(t+p) ge}).

In this case one can see that the strong subadditivity leads to (\ref{x^2le rp}), which is stronger than that derived from Bachas inequality (\ref{r(t+p) ge}). However, this does not mean the strong subadditivity is stronger than Bachas inequality. For example whether the inequality (\ref{r^2-p^2 ge 0}) is derived from the strong subadditivity is a question that the authors are unable to answer at the present time.
 
\subsection{Verification from perturbation}
In this subsection we give a perturbative proof of the strong subadditivity for small deformed Wilson loops, which we proposed at \ref{subsection:F}.

Now let us define loop $C$ as $C=\bigl\{ y \in \mathbf {R}^D |y=x(t) \bigr\} $. From the direct calculation one can see that
second order perturbation of $\log \langle W(C) \rangle$ in $D$ dimensional Yang Mills theory is proportional to
\begin{equation}
w(C)=-\oint_C ds dt \Lambda (r(t,s))\dot{{x}}(t)  \cdot\dot{{x}}(s). 
\end{equation}
where $r(t,s)$ is
\begin{equation}
r(t,s)^i=\left( {x}(t)-{x}(s) \right)^i,
\end{equation}
and $\Lambda (r)$ is a propagator between ${x}(t)$ and ${x}(s)$:
\begin{eqnarray}
 \Lambda(r)&=&\frac{1}{[r^2]^{D/2-1}} (D>2) \label{D(r)}\\
  \Lambda(r)&=&-\log(|r|) \quad(D=2) \label{D(r)2}.
\end{eqnarray}

We introduce small deformed Wilson loop $C+ \delta C=\bigl\{ y \in \mathbf {R}^D |y=x(t)+\delta x(t) \bigr\} $.
Then  we expand the change of $\Lambda(r(t,s))\dot{{x}}(t)  \cdot\dot{{x}}(s)$ with respect to $\delta x(t)$.
To check whether the strong subadditivity is satisfied we consider 
\begin{eqnarray}
\Delta(t,s)&=&\log \frac{\langle W(C+\delta C_1) \rangle \langle W(C+\delta C_2) \rangle}{\langle W(C) \rangle \langle W(C+\delta C_1 +\delta C_2) \rangle}\nonumber \\
&=& w(C+ \delta C_1)+w(C +\delta  C_2)-w(C)-w(C+ \delta C_1 +\delta  C_2).
\end{eqnarray}
Now we consider the case where $\delta x_1$ and $\delta x_2$ expand the original Wilson loop $C$, i.e. are on the outside of $\dot{x}$. Furthermore, we let $\delta x_1$ and $\delta x_2$ do not expand the same point of $C$ i.e. we consider the case where
\begin{equation}
\delta x_1(t) ^i  \delta x_2(t)^j=0 .
\end{equation}
In the end of this section we will consider other cases.

 By expanding $\Delta(t,s)$ by $\delta x(t)$ we have
 \begin{eqnarray}
\Delta(t,s)=
2\bigl( \delta^{il} \delta ^{jm} \delta ^{kn}
+\delta^{kj} \delta^{im}\delta^{ln} -\delta^{kl} \delta^{im} \delta^{jn}
-\delta^{ij}\delta^{km} \delta ^{ln}\bigr )\delta{x}(t)_1^i \delta{x}(s )_2^j \dot{x}(t)^k \dot{x}(s)^l  \Lambda(r)_{mn},
\end{eqnarray}
where $\Lambda_{mn}(r)= \frac{\partial}{\partial r^m}\frac{\partial}{\partial r^n} \Lambda(r)$. 
Here we simplify the equation using convertibility of $t$ and $s$ and use partial integration of $t$ and $s$.

\begin{figure}[htbp]
  \begin{center}
    \includegraphics[scale=1]{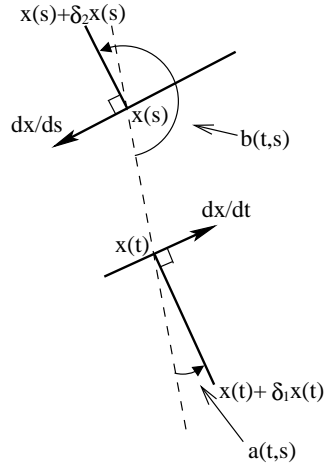}
  \end{center}
  \caption{Configuration of $x(t),x(s),x(t)-x(s),\dot{x}(t),\dot{x}(s),\delta x_1(t)$, and $\delta x_2(s)$}  \label{fig:angle.eps}
\end{figure}

Let angles of $\delta x_1(t)$ and $\delta x_2(s)$ from ${x(t)}-{x(s)}$ be $a(t,s)$ and $b(t,s)$ respectively, then angles of $\dot{x}(t),\dot{x}(s)$ from ${x(t)}-{x(s)}$ are $\pi/2+a,\pi/2+b$, respectively (Fig.\ref{fig:angle.eps}).

Using $a$, $b$, we have
\begin{eqnarray}
\delta x_1(t)\cdot \delta x_2(s)&=&|\delta x_1(t)||\delta x_2(s)|\cos(b-a)\label{76}\\
\delta x_1(t)\cdot \dot{x}(s)&=&|\delta x_1(t)||\dot{x}(s)|\cos(\pi/2+b-a)\label{77}\\
\delta x_2(s)\cdot \dot{x}(t)&=&|\delta x_2(s)||\dot{x}(t)|\cos(\pi/2+a-b)\label{78}\\
\dot{x}(t)\cdot \dot{x}(s)&=& |\dot{x}(t)||\dot{x}(s)|\cos(b-a)\label{79}.
\end{eqnarray}
Using
\begin{equation}
 \frac{\partial}{\partial r^i}\frac{\partial}{\partial r^j} \Lambda(r)=\frac{r^2 \delta^{ij}-r^i r^j}{r^3}\frac{\partial \Lambda}{\partial |r|}  +\frac{r^i r^j}{r^2}  \frac{\partial^2 \Lambda}{\partial |r|^2},
\end{equation} 
and substituting (\ref{D(r)})(\ref{D(r)2})(\ref{76})(\ref{77})(\ref{78})(\ref{79}), we have
\begin{eqnarray}
\Delta(t,s)&=&|\dot{x}(t)||\dot{x}(s)||\delta {x}_1(t)||\delta {x}_2(s)|f(a,b) \nonumber\\ 
f(a,b)&=&-2\frac{1}{r}\frac{\partial \Lambda}{\partial |r|}-2\frac{\partial^2 \Lambda}{\partial |r|^2}=-2\frac{(D-2)^2}{(r^2)^{D/2}}\le 0 \quad (D\ge 2).
\end{eqnarray}
 Thus the inequality:
\begin{equation}
w(C)+w(C+ \delta C_1 +\delta  C_2) \ge w(C+\delta  C_1)+w(x +\delta  C_2)
\end{equation}
is satisfied. This leads to 
\begin{equation}
\langle W(C) \rangle \langle W(C+ \delta C_1 +\delta  C_2)\rangle \ge \langle W(C+\delta  C_1) \rangle \langle W(C +\delta  C_2)\rangle.
\end{equation}

When $D=2$, the inequality saturates. This is because in QCD${}_2$ Wilson loops without crossing obey purely area law $W(C)= e^{-\frac{\lambda }{2} A(C)}$ at leading order of $\lambda$ \cite{Makeenko:2002uj}.

This inequality supports the strong subadditivity.
Because in this case 
\begin{eqnarray}
Int(C+ \delta C_1) \cup Int(C+ \delta C _2) &=&Int(C+ \delta C_1 +\delta C_2) \\
Int(C+ \delta C_1) \cap Int(C+ \delta C_2) &=&Int(C),
\end{eqnarray}
where $Int(C)$ is the interior of the loop $C$.

Thus far, we only considered cases where $\delta x_1$ and $\delta x_2$ both expand the original Wilson loop and $\delta x_1(t) ^i  \delta x_2(t)^j=0$. In general situations by considering loops $C'$, $C'+\delta C_1' $,and $C' +\delta C_2'$ as
\begin{equation}
C'=\partial \bigl(Int (C+\delta C_1) \cap Int (C+\delta C_1)\bigr), \quad C'+\delta C_1' =C+\delta C_1,\quad  C'+\delta C_2' =C+\delta C_2 \label{'1}
\end{equation}
we have 
\begin{equation}
C'+\delta C_1'+\delta C_2'=\partial \bigl(Int (C+\delta C_1) \cup Int (C+\delta C_1)\bigr), \label{'2}
\end{equation}
then we regain original situations.
Using $C'$, $C'+\delta C_1' $, and $C' +\delta C_2'$ we can again prove
\begin{equation}
\langle W(C')\rangle \langle W (C'+\delta C_1'+ \delta C_2') \rangle \ge \langle W(C'+\delta  C_1') \rangle \langle W(C' +\delta  C_2')\rangle,
\end{equation}
 and this is equivalent to the strong subadditivity of $C$, $C+\delta C_1$, and $C+\delta C_2$ since we have (\ref{'1})(\ref{'2}).

\section{Conclusion and Outlook}

In this paper, we proposed the strong subadditivity of Wilson loops, motivated by that of entanglement entropy, and we checked whether it is satisfied in many situations.

Firstly, we checked in the strong coupling region assuming minimal surface conjecture.
Secondly, we checked in the case where Wilson loops are symmetric using Bachas inequality.
Thirdly, we gave a perturbative proof for small-deformed Wilson loops in the weak coupling region.

These results suggest that the strong subadditivity, which has a profound physical meaning, is satisfied in any coupling region for any Wilson loops. Furthermore second and last results give us  new verifications for the minimal surface conjecture.

In this paper, we have omitted the effects of scalar fields, which is necessary to consider AdS/CFT.
Because by deforming the geometry of AdS space one can make scalar fields massive and thus decouple them \cite{Polchinski:2000uf}. 
 One can also prove Bachas inequality for a gauge theory with scalar fields as can be seen in \cite{Dorn:1999yd}.
Therefore, in this paper we neglect the effect of scalar fields.

Many questions remain unsolved in this paper.
The most important one is the proof of the strong subadditivity for arbitrarily-shaped Wilson loops in an arbitrary coupling region. As discussed, the strong subadditivity has a deep connection with that for entanglement entropy and Bachas inequality. Both are derived from the positive definiteness of the Hilbert space. Therefore, we consider that the strong subadditivity of Wilson loops might be also a consequence of it.

In this paper we mainly treat Wilson loops in the same flat plane.
The strong subadditivity for more general cases is also a problem that remains.
In section \ref{sec minimal}, we gave a holographic proof of the strong subadditivity of Wilson loops in the same curved surface. A proof by the gauge theory is an open problem.

\begin{figure}[htbp]
  \begin{center}
    \includegraphics[scale=0.6]{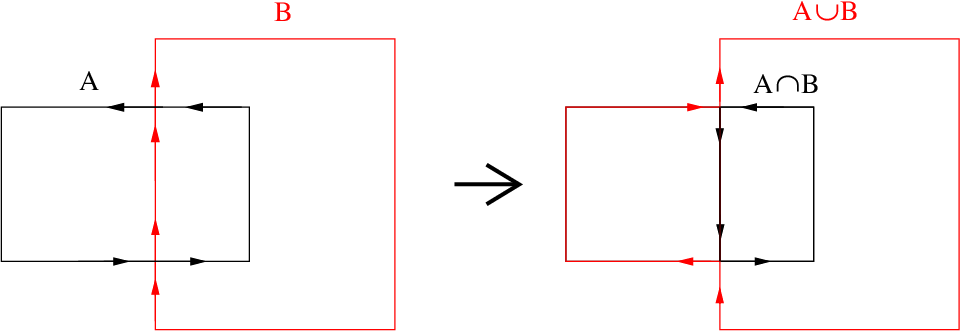}
  \end{center}
  \caption{By changing crossing lines into uncrossing lines, we can change two Wilson loops around $A$ and $B$  into those around $A\cup B$ and $A\cap B$}
  \label{fig:cross.eps}
\end{figure}

As an outlook, we now consider one generalization of the strong subadditivity of Wilson loops. We mention that  for crossing loops in the same surface, by changing crossing loops into uncrossing loops, we can change two Wilson loops around $A$ and $B$  into loops around $A\cup B$ and $A\cap B$ (Fig.\ref{fig:cross.eps}). 

\begin{figure}[htbp]
  \begin{center}
    \includegraphics[scale=0.5]{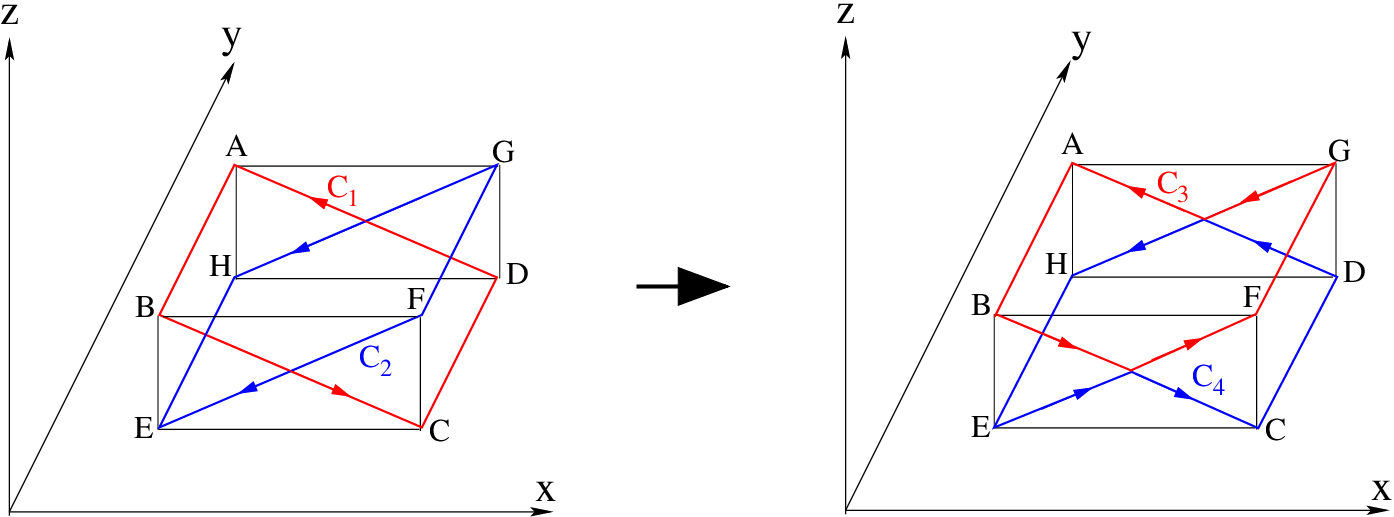}
      \end{center}
  \caption{Two sterically crossing rectangles (left) and their "uncrossed" loops (right).
 In the neighborhood of each crossing point, two loops are locally on the same surface.   }
  \label{fig:hanada.eps}
\end{figure}

Therefore one generalization of the strong subadditivity exists when two loops cross sterically while in the neighborhood of every crossing point $P_i$ there is a surface $S_i$ on which two loops exist. The case shown in the left side of Fig.\ref{fig:cross.eps} is an example of this.
In this instance, one can generalize the strong subadditivity as an inequality between original loops and loops whose crossing points are changed  into uncrossing points (Fig.\ref{fig:hanada.eps}).
\begin{eqnarray}
\langle W(C_1) \rangle \langle W (C_2)\rangle  \le \langle W(C_3) \rangle \langle W (C_4)\rangle, \quad (C_1,C_2 \stackrel{\mathrm{{ uncross}}} \longrightarrow  C_3,C_4)  \label{generalized SS}
\end{eqnarray}

As we stated at the end of section \ref{subsection Cusp}, the generalized the strong subadditivity of sterically-crossing loops (\ref{generalized SS}) is also derived from the convexity of the cusp anomalous dimensions (\ref{conv}). 

\newpage
\begin{center} \end{center}
 {\bf Acknowledgement}:
 We are extremely grateful to  T.Azeyanagi, K.Katayama, T.Nishioka T.Takayanagi, C.Ward  and S.Yamato for the carefully reading out  manuscript and for valuable discussions.
 This research is supported in part by Japan Society for the Promotion of Science Research Fellowships for 
 Young Scientists.

\end{document}